\documentclass[12pt]{article}

\usepackage[margin=1in,nohead,a4paper,dvips]{geometry}
\usepackage{amsfonts}

\def\da{\downarrow}
\def\iu{\mathrm i}
\def\la{\langle}
\def\ra{\rangle}
\def\ua{\uparrow}

\newtheorem{con}{Conjecture}

\begin{document}

\title{Spin chains and combinatorics}
\author{A.~V.~Razumov, Yu.~G.~Stroganov\\
\small \it Institute for High Energy Physics\\[-.5em]
\small \it 142284 Protvino, Moscow region, Russia}
\date{}

\maketitle

\begin{abstract}
In this letter we continue the investigation of finite XXZ spin chains with
periodic boundary conditions and odd number of sites, initiated in paper
\cite{S}. As it turned out, for a special value of the
asymmetry parameter
$\Delta=-1/2$ the Hamiltonian of the system has an eigenvalue, which is
exactly proportional to the number of sites $E=-3N/2$. Using {\sc
Mathematica} we have found explicitly the corresponding eigenvectors for $N
\le 17$. The obtained results support the conjecture of paper \cite{S} that
this special eigenvalue corresponds to the ground state vector. We make a
lot of conjectures concerning the correlations of the model. Many
remarkable relations between the wave function components are noticed. It
is turned out, for example, that the ratio of the largest component to the
least one is equal to the number of the alternating sing matrices. 
\end{abstract}

\leftskip 15em
{\it The Hobbits of the Shire and of Bree had at this time,
for probably a thousand years, adopted the Common Speech.
They used it in their own manner freely and carelessly; 
though the more learned among them had still at their 
command a more formal language when occasion required.}

\hbox to \hsize{\hfil (J.~R.~R.~Tolkien)}

\leftskip 0em

\vskip 1em

The XXZ quantum spin chain model with periodic boundary conditions is one
of the most popular integrable models which has been investigating by the
Bethe Ansatz method during the last 35 years~\cite{Yang}. It is described
by the Hamiltonian
\begin{equation}
H_{XXZ} = -\sum_{j=1}^{N} \left\{ \sigma_j^{x} \sigma_{j+1}^{x}   +
\sigma_j^{y} \sigma_{j+1}^{y} + \Delta \sigma_j^z \sigma_{j+1}^z \right\},
\qquad \vec \sigma_{N+1} = \vec \sigma_{1}.
\label{H1}
\end{equation}
The advantages and shortcomings of the Bethe method are well known and we
do not discuss them here. We do not discuss alternative methods elaborated
by many people as well. We want only to draw attention of experts to some
unique possibility to surpass these methods in one very special case.

As the starting point of our consideration we take the fact that for
$\Delta = -1/2$ and odd number of sites $N = 2n + 1$, Hamiltonian
(\ref{H1}) has an eigenvalue $E = -3N/2$. This fact was proved in paper
\cite{S}. The existence of such eigenvalue was earlier discovered
numerically by Alcaraz, Barber and Batchelor \cite{ABB}. Below we study the properties of the eigenvectors, corresponding
to this special eigenvalue. We use the natural basis where all operators
$\sigma^z_j, \> j=1,\ldots, N$, are diagonal.

For the case $N=3$ we have $2^N = 8$ states and it is not difficult to find
all eigenvalues and eigenvectors. The results of calculations are
described by Table \ref{t1}, where $\omega = \exp (2 \pi \iu / 3)$.
\begin{table}[ht]
\caption{All eigenvectors for the case $N = 3$}
{\small
\[
\begin{array}{lllrr}
\hline \hline
\multicolumn{3}{c}{\vrule height 1.1em width 0pt |\Psi \ra} &
\multicolumn{1}{c}{S_z} &
\multicolumn{1}{c}{E} \\
\hline
\vrule height 1.1em width 0pt
\mid \da \da \da \ra & & &
  -3/2 & 3/2 \\[.3em]
\mid \ua \da \da \ra + &
\hspace{-.7em} \mid \da \ua \da \ra + &
\hspace{-.7em} \mid \da \da \ua \ra  &
  -1/2 & -9/2 \\[.3em]
\mid \ua \da \da \ra + \omega &
\hspace{-.7em} \mid \da \ua \da \ra + \omega ^2 &
\hspace{-.7em} \mid \da \da \ua \ra &
  -1/2 & 3/2 \\[.3em]
\mid \ua \da \da \ra + \omega^2 &
\hspace{-.7em} \mid \da \ua \da \ra + \omega &
\hspace{-.7em} \mid \da \da \ua \ra &
  -1/2 & 3/2 \\[.3em]
\mid \da \ua \ua \ra + &
\hspace{-.7em} \mid \ua \da \ua \ra + &
\hspace{-.7em} \mid \ua \ua \da \ra &
  1/2 & -9/2 \\[.3em]
\mid \da \ua \ua \ra + \omega &
\hspace{-.7em} \mid \ua \da \ua \ra + \omega^2 &
\hspace{-.7em} \mid \ua \ua \da \ra  &
  1/2 & 3/2 \\[.3em]
\mid \da \ua \ua \ra + \omega^2 &
\hspace{-.7em} \mid \ua \da \ua \ra + \omega &
\hspace{-.7em} \mid \ua \ua \da \ra  &
  1/2 & 3/2 \\[.3em]
\mid \ua \ua \ua \ra & & & 
  3/2 & 3/2 \\[.3em]
\hline
\end{array}
\]}
\label{t1}
\end{table}
Already this simplest case demonstrates some characteristic properties of
the spectrum of Hamiltonian (\ref{H1}) and of its eigenvectors. First note
that the operator
\[
R = \prod^{N}_{j=1} \sigma^x_j,
\]
reversing the $z$-axis projections of all spins, commutes with Hamiltonian
(\ref{H1}), hence, every eigenvector has a partner with the same energy
and the opposite value of $S_z$. In particular, the ground state is
two-fold degenerate. 

We see that our special eigenvalue corresponds to the ground state of the
system. Then it is not surprising that it is unique for a fixed value of
$S_z$ and shift invariant. It is natural to start to suspect that this fact
in not accidental and that our special eigenvalue corresponds to the ground
state for an arbitrary odd $N$. Another indication supporting this
hypothesis is that as was shown in paper \cite{S}, the eigenvalue $-3N/2$
occurs in the sectors with $S_z = \pm 1/2$ that resembles the value
$S_z = 0$ occurring for the ground state in the case of an
even $N$~\cite{L}.

Let us proceed to larger values of $N$. Due to the symmetry described
above it suffices to consider only positive or only negative values of
$S_z$. For definiteness let us restrict ourselves to the case $S_z < 0$.
Moreover, having in mind our hypothesis, we will consider only the shift
invariant states.

The shift invariant eigenvectors for $N = 5$ and $S_z < 0$ are
given in Table~\ref{t2}.
\begin{table}[ht]
\caption{All shift invariant eigenvectors for the case $N = 5$ and
$S_z < 0$}
{\small
\[
\begin{array}{lllrr}
\hline \hline
\multicolumn{3}{c}{\vrule height 1.1em width 0pt | \Psi \ra} &
\multicolumn{1}{c}{S_z} &
\multicolumn{1}{c}{E} \\
\hline
\vrule height 1.3em width 0pt &
\hspace{-.7em} \overline{\mid \da \da \da \da \da \ra} & & -5/2 & 5/2
\\[.3em]
& \hspace{-.7em} \overline{\mid \da \da \da \da \ua \ra} & & -3/2 & -3/2
\\[.3em]
2 & \hspace{-.7em} \overline{\mid \da \da \da \ua \ua \ra} - & 
\hspace{-.7em} \overline{\mid \da \da \ua \da \ua \ra} & -1/2 & 5/2
\\[.3em]
& \hspace{-.7em} \overline{\mid \da \da \da \ua \ua \ra} + 2 & 
\hspace{-.7em} \overline{\mid \da \da \ua \da \ua \ra} & -1/2 & -15/2
\\[.3em]
\hline
\end{array}
\]}
\label{t2}
\end{table}
A line above a vector means that we use the corresponding shift invariant
combinations, for example,
\[
\overline{\mid \da \da \ua \da \ua \ra} = 
\mid \da \da \ua \da \ua \ra + 
\mid \da \ua \da \ua \da \ra +
\mid \ua \da \ua \da \da \ra +
\mid \da \ua \da \da \ua \ra +
\mid \ua \da \da \ua \da \ra.
\]
The coincidence of the energy of the state with $S_z = - 5/2$ and the
energy of some state with $S_z = - 1/2$ can be explained by the loop
symmetry discovered by Korepanov~\cite{Korepanov} and recently 
elaborated by Deguchi, Fabricius and McCoy~\cite{Mc}.

Table \ref{t3} describes some shift invariant eigenvectors for the
case $N = 7$.
\begin{table}[ht]
\caption{Some shift invariant eigenvectors for the case $N = 7$
and $S_z < 0$}
{\small
\[
\begin{array}{llrr}
\hline \hline
\multicolumn{2}{c}{\vrule height 1.1em width 0pt | \Psi \ra} &
\multicolumn{1}{c}{S_z} &
\multicolumn{1}{c}{E} \\
\hline
\vline height 1.3em width 0pt & \hspace{-.7em} 
\overline{\mid \da \da \da \da \da \da \da \ra} &
  -7/2 & 7/2 \\[.3em]
& \hspace{-.7em} 
\overline{\mid \da \da \da \da \da \da \ua \ra} &
  -5/2 & -5/2 \\[.3em]
2 & \hspace{-.7em} 
\overline{\mid \da \da \da \da \ua \ua \ua \ra} -
\overline{\mid \da \da \da \ua \ua \da \ua \ra} +
\overline{\mid \da \da \ua \da \da \ua \ua \ra} -
\overline{\mid \da \da \da \ua \da \ua \ua \ra} & 
  -1/2 & 7/2 \\[.3em]
& \hspace{-.7em} 
\overline{\mid \da \da \da \da \ua \ua \ua \ra} - 2 
\overline{\mid \da \da \ua \da \da \ua \ua \ra} +
\overline{\mid \da \da \ua \da \ua \da \ua \ra} & 
  -1/2 & 3/2 \\[.3em]
& \hspace{-.7em} 
\overline{\mid \da \da \da \ua \ua \da \ua \ra} -
\overline{\mid \da \da \da \ua \da \ua \ua \ra} 
  & -1/2 & 3/2 \\[.3em]
& \hspace{-.7em} 
\overline{\mid \da \da \da \da \ua \ua \ua \ra} +
\overline{\mid \da \da \da \ua \ua \da \ua \ra} +
\overline{\mid \da \da \da \ua \da \ua \ua \ra} -
\overline{\mid \da \da \ua \da \ua \da \ua \ra} & 
  -1/2 & -5/2 \\[.3em]
& \hspace{-.7em} 
\overline{\mid \da \da \da \da \ua \ua \ua \ra} + 3
\overline{\mid \da \da \da \ua \ua \da \ua \ra} + 4
\overline{\mid \da \da \ua \da \da \ua \ua \ra} + 3
\overline{\mid \da \da \da \ua \da \ua \ua \ra} + 7
\overline{\mid \da \da \ua \da \ua \da \ua \ra} & 
  -1/2 & -21/2 \\[.3em]
\hline
\end{array}
\]}
\label{t3}
\end{table}
Actually we did not include there only the eigenvectors belonging to the
sector with $S_z = -3/2$. The energy values in this sector are determined
by the third order equation
\[
8 E^3 + 28 E^2 - 298 E - 411 = 0
\]
which has no rational roots. The approximate values of the roots are $E
\approx -7.5$, $E \approx -1.3$ and $E \approx 5.3$. 

It is curious that the five states belonging to the sector $S_z = -1/2$ 
which for an arbitrary asymmetry parameter $\Delta$ lead to an equation of
the fifth degree, have halfinteger energies for $\Delta=-1/2$.  It is a
manifestation of the general rule that in the case of $\Delta = -1/2$ the
three spin wave energies for the states with zero momentum belong to the
field $\mathbb Q$ of rationals extended with the root of unity $\exp(2 \pi
\iu/(N-1))$. It will be, probably, published elsewhere but now
we are interested in the ground state only.

We see that for $N = 5$ and $N = 7$ the special energy eigenvalue again
corresponds to the ground state. So we confirm the following conjecture,
formulated in paper \cite{S} and supported by numerical calculations
by Alcaraz, Barber and Batchelor~\cite{ABB}:
\begin{con}
The ground states of Hamiltonian {\rm (\ref{H1})} for an arbitrary odd $N$
have the energy $E = -3N / 2$ and $S_z = \pm 1/2$.
\end{con}
If the above conjecture is true, then we have a real possibility to study
the correlations for the finite periodic XXZ chains with $\Delta=-1/2$ and
odd number of cites. Below we report a few conjectures related to the
average over the states corresponding to the eigenvalue $-3N/2$.

Let us summarize the information related to the ground state for $N=3, 5,
7$. We mark the components by the eigenvalues of the operators
\[
a_j = (1+\sigma^z_{j})/2.
\]
The nonzero wave function components are
\[
\begin{array}{ll}
N=3: & \psi_{001} = 1; \\
N=5: & \psi_{00011} = 1, \quad \psi_{00101} = 2; \\
N=7: & \psi_{0000111} = 1, \quad \psi_{0001101} = \psi_{0001011} = 3, \quad
\psi_{0010011} = 4 \quad \psi_{0010101} = 7.
\end{array}
\]
All components not included in the list can be obtained by shifting.
Notice that the components of the ground state are positive in accordance
with the Perron--Frobenius theorem.

Let us continue the list. For $N=9$ the components of the eigenvector with
the energy~$-27/2$ and $S_z = -1/2$ are
\[
\begin{array}{llll}
\psi_{000001111} = 1, \qquad &
\psi_{000010111} = 4, \qquad & 
\psi_{000011011} = 6, \qquad &
\psi_{000100111} = 7, \\
\psi_{000101011} = 17, \qquad &
\psi_{000101101} = 14, \qquad &
\psi_{000110011} = 12, \qquad &
\psi_{001001011} = 21, \\ 
& \psi_{001010011} = 25, \qquad & 
\psi_{001010101} = 42.
\nonumber
\end{array}
\]
We omit nonzero components which can be obtained by the reflection
of the order of sites since this transformation is a symmetry 
of our state, as it is for the ground state. For example, we have
\[
\psi_{000011101} = \psi_{000010111} = 4.
\]

It is incredible at the turn of the millennium but up to this place the
Computer was switched off! But the linear system for $N = 11$ contains 42
equation, so we invited the {\sc Mathematica} and obtained:
\[
\begin{array}{llll}
\psi_{00000011111} = 1, \qquad &
\psi_{00000101111} = 5, \qquad &
\psi_{00000110111} = 10, \qquad &
\psi_{00001001111} = 11, \\
\psi_{00001010111} = 34, &
\psi_{00001011011} = 41, &
\psi_{00001011101} = 23, &
\psi_{00001100111} = 30, \\
\psi_{00001101011} = 60, &
\psi_{00010001111} = 14, &
\psi_{00010010111} = 52, &
\psi_{00010011011} = 73, \\
\psi_{00010011101} = 46, &
\psi_{00010100111} = 75, &
\psi_{00010101011} = 169, &
\psi_{00010101101} = 128, \\
\psi_{00010110011} = 101, &
\psi_{00011000111} = 42, &
\psi_{00011001011} = 114, &
\psi_{00100100111} = 81, \\
\psi_{00100101011} = 203, &
\psi_{00100101101} = 174, &
\psi_{00100110011} = 141, &
\psi_{00101001011} = 226, \\
& \psi_{00101010011} = 260, &
\psi_{00101010101} = 429. &
\end{array}
\]
It was a turning point of our work. The problem assumed a combinatorial
character. The number 429 reminded us about the
history of ASM conjecture~\cite{BP}. The number of alternating sign $n
\times n$ matrices is given by formula
\begin{equation}
\label{An}
A_n=\prod_{j=0}^{n-1}\frac{(3j+1)!}{(n+j)!}.
\end{equation}
The sequence $A_n$ goes as
\[
1,\ 2,\ 7,\ 42,\ 429,\ 7436,\ 218348, \ 10850216, \ldots
\]
The largest component of the wave functions for $N=3$, 5, 7, 9, 11
goes in the same way! {\sc Mathematica} helped us to find components for $N
= 13$, 15 and 17. For $N = 19$ the Computer was immersed in its
calculations 
for too long, and we have lost patience. All the obtained results confirm
\begin{con}
If we chose the normalization of the state vector under consideration so
that $\psi_{0 \ldots 01 \ldots 1} = 1$ then all other components are
integer and the largest
component is given by $\psi_{0010101 \ldots 01} = A_n$.
\end{con}

Moreover we obtained that the components of the state vector are positive.
Therefore, in accordance with the Perron--Frobenius theorem the special
energy eigenvalue corresponds to the ground state at least in the sectors
with $S_z = \pm 1/2$.

Now we are analyzing the obtained information and can already formulate
some conjectures which are verified for odd $N \le 17$.

For calculations of correlations one needs to know the norm of the ground
state. For $N = 2n+1$ we denote it by $\mathcal N_n$. The results of
calculations looks as
\[
{\mathcal N}_1 = \sqrt{3}, \quad {\mathcal N}_2 = 5, \quad {\mathcal N}_3 =
14 \sqrt{3}, \quad {\mathcal N}_4 = 198, \quad \ldots
\]
They are in agreement with
\begin{con}
If we chose the normalization so that $\psi_{0\ldots 01 \ldots 1} = 1$ the
norm of our state is
\[
{\cal N}_n = \frac{\sqrt{3^n}}{2^n} \, \frac{ 2 \cdot 5
\dots (3n-1)} {1 \cdot 3 \dots (2n-1)}\  A_n.
\]
\end{con}

It is curious that the sequence of the linear sums of the components also
display the simple parametrization.
\begin{con} The sum of all components of the state vector under
consideration is equal to $(\sqrt{3})^n{\mathcal N}_n$.
\end{con}

As was shown in paper \cite{S}, the simplest nontrivial correlations 
are described by the formula
\[
\la \sigma^z_j \sigma^z_{j+1} \ra = -\frac{1}{2}+\frac{3}{2(2n+1)^2}.
\]
This implies that
\[
\la a_j \, a_{j+1} \ra = \frac{(n-1)n}{2(2n+1)^2}.
\]
Our results are in agreement with this formula.

We have also a fit for the next 2-spin correlations:
\begin{con}
\[
\la a_j \, a_{j+2} \ra = \frac{(n-1)}{4 (2n-1)(2n+1)^2 (2n+3)}
\left[ \frac{71}{4}n^3+\frac{149}{4}n^2+18n+9 \right].
\]
\end{con}
In spite of its ugly look the above formula possesses some hidden symmetry,
as well as the next one.
\begin{con}
There is a simple formula for the correlations called the Probabilities of
Formation of Ferromagnetic String {\rm\cite{KIB}}:
\[
\frac{\la a_1\>a_2\>\dots\>a_{k-1} \ra}{\la a_1\>a_2\>\dots\>a_{k} \ra}
=\frac{(2k-2)!\>(2k-1)!\>(2n+k)!\>(n-k)!}{(k-1)!\>(3k-2)!\>(2n-k+1)!\>
(n+k-1)!}.
\]
\end{con}
If this conjecture is correct then we obtain in the thermodynamic 
limit
$$
\la a_1\>a_2\>\dots\>a_{k} \ra =
\biggl(\frac{\sqrt{3}}{2}\biggr)^{3 k^2}\ \prod_{m=1}^k
\frac{\Gamma(m-1/3)\Gamma(m+1/3)}{\Gamma(m-1/2)\Gamma(m+1/2)}.
$$
Using the standard properties of $\Gamma(x)$ we find an asymptotic
of this correlator for large values of $k$
$$
\la a_1\>a_2\>\dots\>a_{k} \ra \approx
c \, \biggl(\frac{\sqrt{3}}{2}\biggr)^{3 k^2}\ 
k^{-\frac{5}{36}},
$$
where 
$$
c = \exp \left[\int_0^{\infty}\left(\frac{5}{36} \exp(-t)-
\frac{\sinh (5t/12) \sinh (t/12)}{\sinh ^2 (t/2)}\right)
\frac{dt}{t} \right] \approx 0.77466966.
$$

The last conjecture touches the both ground states. The additional
transverse magnetic field which is described by
\[
\Delta H = h\ \sum_{j=1}^N \sigma^x_j,
\]
splits the energies of the two ground states. In the first order the
splitting is determined by the matrix element
\[
M = \la S_z=1/2|\sigma^x_j|S_z=-1/2 \ra.
\]
Our data are in agreement with 
\begin{con}
\[
(2n+1)\ M =\frac{4\cdot 7\cdot\dots\>(3n+1)}
{2\cdot 5\cdot\dots\>(3n-1)}.
\]
\end{con}

In conclusion we would like to concentrate the reader's attention on the
appearance in our investigation of many combinatorial relations and
remarkable combinatorial numbers. We consider it as an indication on the
existence of rather simple expressions for the components of the ground
state eigenvector of the model.

{\it Acknowledgments} The authors would like to thank M.~T.~Batchelor,
V.~E.~Korepin, I.~G.~Korepanov, B.~M.~McCoy, R.~I.~Nepomechie 
and M.~S.~Plyushchay for
their interest in our work and comments.
The work was supported in part by the Russian
Foundation for Basic Research under grant \# 98--01--00015 (A.V.R) and
grant \# 98--01--00070 (Yu.G.S).

\end{document}